\documentstyle[12pt,preprint]{aastex}

\begin{document}




\shortauthors{Crutcher, et al.} 
\shorttitle{Magnetic Fields in Dark Clouds}

\title{SCUBA Polarization Measurements of the Magnetic Field Strengths in the L183, L1544, and L43 Prestellar Cores}

\author{Richard M. Crutcher\altaffilmark{1}}

\affil{Department of Astronomy, University of Illinois, 1002 W. Green Street, Urbana, IL 61801, USA }

\author{D. J. Nutter\altaffilmark{2}, D. Ward-Thompson\altaffilmark{3} }
\affil{ Department of Physics \& Astronomy, 5, The Parade, Cardiff University, Cardiff CF2 3YB, UK }

\author{J. M. Kirk\altaffilmark{4}}

\affil{Department of Astronomy, University of Illinois, 1002 W. Green Street, Urbana, IL 61801, USA }

\altaffiltext{1}{crutcher@uiuc.edu}
\altaffiltext{2}{David.Nutter@astro.cf.ac.uk}
\altaffiltext{3}{Derek.Ward-Thompson@astro.cf.ac.uk}
\altaffiltext{4}{jmkirk@astro.uiuc.edu}



\begin{abstract}

We have mapped linearly polarized dust emission from L183 with the JCMT SCUBA polarimeter and have analyzed these and our previously published data for the prestellar cores L183, L1544, and L43 in order to estimate magnetic field strengths in the plane of the sky, $B_{pos}$. The analysis used the Chandrasekhar-Fermi technique, which relates the dispersion in polarization position angles to $B_{pos}$. We have used these estimates of the field strengths (neglecting the unmeasured line-of-sight component) to find the mass-to-magnetic flux ratios $\lambda$ (in units of the critical ratio for magnetic support). Results are $B_{pos} \approx 80$ $\mu$G and $\lambda \approx 2.6$ for L183, $B_{pos} \approx 140$ $\mu$G and $\lambda \approx 2.3$ for L1544, and $B_{pos} \approx 160$ $\mu$G and $\lambda \approx 1.9$ for L43. Hence, without correction for geometrical biases, for all three cores the mass-to-flux ratios are supercritical by a factor of $\sim 2$, and magnetic support cannot prevent collapse. However, a statistical mean correction for geometrical bias may be up to a factor of three; this correction would reduce the individual $\lambda$'s to $\lambda_{cor} \approx 0.9$, 0.8, and 0.6, respectively; these values are approximately critical or slightly subcritical. These data are consistent with models of star formation driven by ambipolar diffusion in a weakly turbulent medium, but cannot rule out models of star formation driven by turbulence. 

\end{abstract}



\section{Introduction}

Understanding star formation is one of the outstanding challenges of modern astrophysics. However, in spite of significant progress in recent years, there remain unanswered fundamental questions about the basic physics of star formation. It remains unclear whether molecular clouds have lifetimes significantly longer than their free-fall collapse times $\tau_{ff}$, typically $\sim 10^6$ years (\citet{hpb01}, \citet{ps02}). If self-gravitating clouds and cores have long lifetimes ($>>\tau_{ff}$), they must be supported against gravitational collapse. One view is that magnetic fields provide this support. Detailed theoretical work on the evolution of clouds with initially magnetically subcritical masses (i.e., fields strong enough to support the cloud), with ambipolar diffusion driving cloud evolution and star formation, have been carried out by a number of groups (e.g., \citet{mc99}). The other extreme from the magnetically dominated star formation point of view is that molecular clouds are intermittent phenomena in an interstellar medium dominated by turbulence, and the problem of cloud support for long time periods is irrelevant (e.g., \citet{e00}). However, even if magnetic fields turn out to be too weak to provide cloud support, they appear to be essential for star formation in any case. A star-forming cloud must lose most of its angular momentum to its surroundings in order to collapse. Outward angular momentum transport by Alfv\'{e}n waves seems to be the only mechanism available for this process \citep{mp86}.

Hence, theoretical studies suggest that magnetic fields may play an important or even crucial role in the evolution of interstellar clouds and in the formation of stars. But the issue is far from settled, on either observational or theoretical grounds. The critical value for the mass that can be supported by a magnetic flux $\Phi$ is $M_{B crit} = \Phi/2\pi\sqrt{G}$ \citep{nn78}; the precise value of the numerical coefficient is slightly model dependant (e.g., \citet{ms76}; \citet{mzgh93}). The crucial parameter is then the ratio of the mass to the magnetic flux, $M/\Phi$; observational determination of this parameter will answer the question whether magnetic fields support clouds and ambipolar diffusion governs the subsequent collapse of clouds to form stars.

We can state $M/\Phi$ in units of the critical value and define $\lambda \equiv (M/\Phi)_{actual}/ (M/\Phi)_{crit}$. $\lambda$ is generally either an assumed or a free parameter in theoretical models of star formation; it must be determined empirically. Inferring $\lambda$ from observations is possible if the column density $N$ and the magnetic field strength $B$ are measured:
\begin{equation}
\lambda = \frac{(M/\Phi)_{observed}}{(M/\Phi)_{crit}} = \frac{m N A/B A}{1/2\pi \sqrt{G}} = 7.6 \times 10^{-21} N(H_2)/B
\end{equation}
where $m = 2.8m_H$ allowing for 10\% He by number, with $N(H_2)$ in cm$^{-2}$ and $B$ in $\mu$G. Note that the area A over which $M$ and $\Phi$ are measured cancels, although of course $B$ and $N$ must be measured over the indentical area of a cloud.

Until recently, measurement of magnetic field strengths by observations of the Zeeman effect in potential star formation regions has been the only method by which $\lambda$ has been determined. \citet{cr99} analyzed all Zeeman data available for molecular cloud cores in order to infer the mean value $\bar{\lambda}$ in molecular clouds. He found that $\bar{\lambda} \sim 2$, indicating that cores are slightly supercritical, i.e., their magnetic fields are too weak for support against gravity by magnetic pressure alone. However, Zeeman detections are rather sparse, particularly in prestellar \citep{wt94} cores. Hence, it is crucially important that a more statistically valid sample of measurements of magnetic field strengths in molecular cores be obtained.

Although observations of the polarization of emission from dust grains do not directly yield the magnitude of the magnetic field, in the early days of interstellar polarization studies \citet{cf53} suggested that analysis of the small-scale randomness of magnetic field lines could yield estimates of the field strengths. The method depends on the fact that turbulent motions will lead to non-uniform magnetic fields (since under interstellar conditions fields will be frozen into the matter). There will therefore be a perturbed or MHD-wave component to the field that should show up as an irregular scatter in polarization positions angles relative to those that would be produced by a regular magnetic field. They showed that the magnitude of that irregularity of field lines could yield the regular field strength in the plane of the sky: 
\begin{equation}
B_{pos} = Q\sqrt{4\pi \rho} \; \frac{\delta V}{\delta \phi} \approx 9.3 \frac{\sqrt{n(H_2)} \;\Delta V}{\delta \phi} \mu G, 
\end{equation}
where $\rho = m n(H_2)$ is the gas density, $\delta V$ is the velocity dispersion, $\delta \phi$ is the dispersion in polarization position angles (in degrees on the right-hand side), $Q$ is a factor of order unity, $n(H_2)$ is the molecular hydrogen density in molecules cm$^{-3}$, and $\Delta V = \sqrt{8 ln2} \; \delta V$ is the FWHM line width in km~s$^{-1}$. 

It is important to know how well the indirect Chandrasekhar-Fermi method works in estimating magnetic field strengths. In order study this question, \citet{osg01}, \citet{p01}, and \citet{hz01} simulated magnetized molecular regions and ``observed'' those simulations in order to compute dust polarization images. They then compared $B_{pos}$ inferred from the ``observed'' dispersions in polarization position angles with the known magnetic field strengths. With no smoothing of the simulations and relatively strong field strengths, such that $\delta \phi < 25^\circ$, \citet{osg01} found $Q \approx 0.46 - 0.51$, while \citet{p01} and \citet{hz01} found $Q \approx 0.3 - 0.4$. The three studies concluded that the Chandrasekhar-Fermi method would yield reliable values of $B_{pos}$ for regions of strong fields (such as molecular clouds) so long as the $Q$ correction was applied. \citet{hz01} then investigated the result of inadequate spatial resolution in observations by smoothing their simulated polarization images and inferring $Q$; with the maximum smoothing considered, $Q \approx 0.1$; i.e., a severe overestimation of $B_{pos}$. However, the simulations did not have sufficient resolution to resolve dense cores sufficiently such that smoothing studies for the dense cores could be reliably carried out \citep{h03}. The \citet{hz01} simulations that were used for the resolution studies only had $128^3$ computational cells.  High-density filaments and cores within those filaments typically had only a small number of computational pixels across them. As stated by \citet{hz01}, ``our simulations refer to a whole molecular cloud region comprising dense filaments and cores as well as the dilute gas.'' Self-gravitating cores are not properly resolved in the simulations, and indeed the simulations were halted when dense filaments formed due to insufficient resolution to follow the evolution further. Hence, the \emph{unsmoothed} simulations -- particularly the higher resolution ($256^3$) ones of \citet{osg01} best represent the data reported here. We therefore adopt their value $Q = 0.5$ in Equation 2. 

The uncertainty in values of $B_{pos}$ determined from equation (2) may be estimated knowing the uncertainties in the quantities on the right side of the equation. First considering only the observational uncertainties, the error is dominated by uncertainties in $n(H_2)$. Both $\Delta V$ and $\delta \phi$ are generally known within $\sim10\%$, but the uncertainty in $n(H_2)$ may be as much as a factor of two, mitigated by the fact that $n(H_2)$ enters only as the square root. Adopting these uncertainties, the uncertainty $\sigma(B)$ in $B_{pos}$ is $\sigma(B)/B_{pos} \approx 0.5$. Systematic uncertainties may be parameterized by the uncertainty in $Q$. Equation (2) requires the velocity dispersion in the plane of the sky, yet what is available observationally is the velocity dispersion along the line of sight; the assumption is made that (at least statistically) turbulent velocities are isotropic. $Q$ includes a mean correction for the fact that several turbulent cells or MHD wavelengths will be averaged through the cloud, reducing $\delta \phi$. Moreover, if the angular resolution of the observations is insufficient to resolve MHD wavelengths, there will be spatial smoothing in the plane of the sky. On the other hand, large-scale structure in the regular field that is not subtracted out before $\delta \phi$ is calculated from the polarization position angles will increase the measured $\delta \phi$. The simulations discussed above suggest that the uncertainty in $Q$ may be less than $\sim 30\%$. For example, \citet{osg01} found $Q$ in the range 0.46 - 0.51 when $\delta \phi < 25^\circ$, which is certainly the case for the three cores considered in this paper, and concluded that ``...the Chandrasekhar-Fermi formula multiplied by a factor ~0.5 yields a good estimate of the plane-of sky magnetic field strength, provided the dispersion in polarization angles is $<25^\circ$...''. Including a 30\% uncertainty in $Q$ gives $\sigma(B)/B_{pos} \approx 0.6$. 

Also, this method gives only two of the three components of {\bf B}; the line-of-sight component is not constrained. Since $B_{pos} = B_{total}\sin\theta$, the statistical average value of $\overline{B}_{pos}$ is
\begin{equation}
\overline{B}_{pos} = \int_{0}^{\pi/2} [\vert{\bf B}\vert \sin \theta] \sin \theta d\theta = \frac{\pi}{4}\vert{\bf B}\vert.
\end{equation}
Although there is not a large underestimate in the total B in using only the values of $B_{pos}$, there can be a large geometrical bias in the directly inferred values of $\lambda$. If {\bf B} is strong, clouds will have an oblate spheroid or disk morphology, with \textbf{B} along the minor axis. To properly measure $\lambda$, one needs $B$ and $N$ along a flux tube, i.e., parallel to the minor axis. But in general $N$ (and $M$) will be overestimated by $1/\cos \theta$ if the sight line has an angle $\theta$ to the minor axis, while $\vert{\bf B}\vert$ (and $\Phi$) is underestimated by $\sin \theta$. Again statistically, 
\begin{equation}
\overline{M/\Phi} = \int_{0}^{\pi/2} \frac{M\cos\theta}{\Phi/\sin\theta} \sin \theta d\theta = \frac{1}{3}(M/\Phi)_{obs}. 
\end{equation}

Thus, mapping of dust polarization provides a new technique for estimating magnetic field strengths and hence $\lambda$. In this paper we report on new SCUBA polarimeter observations of the L183 prestellar core and the analysis of these and previously published observations of the L1544 and L43 prestellar cores \citep{dwt00} to infer the strength of the magnetic fields in these cores. We then find $\lambda$ for the three prestellar cores and discuss the results.

\section{Observations}

Submillimetre continuum polarimetry observations at $\lambda850$ $\mu$m were carried out using the SCUBA (Submillimetre Common User Bolometer Array) camera on the James Clerk Maxwell Telescope (JCMT\footnote{JCMT is operated by the Joint Astronomy Center, Hawaii, on behalf of the UK PPARC, the Netherlands NWO, and the Canadian NRC. SCUBA was built at the Royal Observatory, Edinburgh. SCUBAPOL was built at Queen Mary \& Westfield College, London.}) on Mauna Kea, Hawaii, on the mornings of 1999 March 15 and September 15, as well as 2002 February 15 through 17 from HST 01:30 to 09:30 (UT 11:30 to 19:30). Pointing and focussing were checked using the bright sources Uranus, 1514-240 and IRAS16293, and the pointing was found to be good to $\sim$1--2$^{\prime\prime}$ throughout. All three cores were observed in 1999; in 2002 only L183 was observed.

SCUBA \citep{h99} was used in conjunction with the polarimeter SCUBAPOL \citep{g02} in the 16-position jiggling mode to produce a fully sampled $2.3^\prime$ image \citep{h99}. The observations were carried out while using the secondary mirror to chop $120^{\prime\prime}$ in azimuth at around 7 Hz and synchronously to detect the signal, thus rejecting ``sky'' emission. The method of observing used was to make a full 16-point jiggle map (to produce a Nyquist sampled map), with an integration time of 1 second per point, in each of the ``left'' and ``right'' beams of the telescope (the two beams are produced by the chopping secondary mirror).

This process was repeated at each position of the polarimeter half-wave plate. Then the half-wave plate was rotated to the next position, in steps of 22.5$^\circ$. 16 such positions thus constitute a complete revolution of the half-wave plate, representing 512 seconds of on-source integration, which took about 12 minutes, including overheads \citep{g99, g02}. This was then repeated several times, the data stacked and a sinusoidal curve was fitted to the data to calculate the magnitude and direction of the polarization.

The instrumental polarization of each bolometer was measured on the planets Uranus and Saturn and subtracted from the data before calculating the true source polarization. The mean instrumental polarization was found to be 0.93$\pm$0.27\%. The observations were repeated with a slight offset in each case (caused by sky rotation) so that any individual bolometers with significantly above average noise could be removed without leaving areas of the map with no data. The atmospheric opacity at 225~GHz was monitored by the radiometer located at the Caltech Submillimeter Observatory. The opacity at 225~GHz was 0.06--0.07 during the 1999 observations, and 0.04--0.07 during the 2002 observations, typical of good conditions at the site.

Each individual night was seen to be stable within itself, which is important for the polarimetry measurements. The data from each night were reduced separately and found to produce consistent results. Data from all nights were then co-added to produce a higher signal-to-noise ratio final map. We have previously published the data from the 1999 observing run (Ward-Thompson et al. 2000). The data we publish herein are consistent with the previous paper, but have a higher signal-to-noise ratio, and hence more polarization vectors at lower intensity levels. 

\section{Results}

Figure 1 shows our map for the prestellar core L183. Intensities are shown as a grey-scale image with contours overlaid; the vectors represent the polarized intensity (PI) measured at each position. They have been rotated by 90$^\circ$ to show the inferred direction of the magnetic field. In order to improve the signal-to-noise ratio for the map of PI, it was smoothed from the telescope resolution of $14^{\prime\prime}$ to $21^{\prime\prime}$ resolution; the intensity map was not smoothed. The typical PI sensitivity is  $s_p \approx 2$ mJy beam$^{-1}$, although it varies from $\sim$ 1.5 mJy beam$^{-1}$ near the intensity peak to $\sim$ 4 mJy beam$^{-1}$ in the low intensity regions. This variation is mainly due to the fact that at positions with low total intensity the fitting process ($\S2$) is more affected by atmospheric instabilities than it is at high intensity positions. 

Our goal is to infer magnetic field strengths and mass-to-magnetic flux ratios from equations (1) and (2). For this we need estimates of the dispersion in polarization position angles, volume density, column density, and velocity dispersion for the same region of each core. In each subsection below we discuss details of the measurements of the needed parameters. We first discuss here factors common to all three cores. 

For each position at which we have a significant polarization measurement ($PI/s_p > 2$), we have the Stokes Q and U values and their associated measurement uncertainties. From these we compute the position angle $\phi$ and its measurement uncertainty $s_\phi$  for each position. We weight each measurement by $s_\phi^{-2}$ and compute the weighted mean position angle $\overline{\phi}$ for each core. In the absence of measurement uncertainty, the deviation of each measured $\phi$ from $\overline{\phi}$ would be due to the random magnetic field component at that position; we compute the dispersion in position angles $\delta \phi$ from these deviations, with each $\phi - \bar{\phi}$ weighted by $s_\phi^{-2}$. From the measurement uncertainties $s_\phi$ we also estimate $\sigma_\phi$, the dispersion in the position angles expected just from measurement uncertainty with no intrinsic scatter in position angles. In the absence of any random component of the magnetic field, $\delta \phi = \sigma_\phi$. We find that typically $\delta \phi \approx 3 \sigma_\phi$, which means that we have measured a random magnetic field. We correct the upward bias of $\delta \phi$ due to $\sigma_\phi$ (such that $\delta\phi_{actual}^{2}=\delta\phi_{meas}^{2}-\sigma_{\phi}^{2}$) and use this corrected $\delta \phi$ in equation (2). It should be noted that the ratio $\delta \phi/\sigma_\phi$ is not a measure of the uncertainty in our determination  of $\delta \phi$. $\sigma_\phi$ represents the mean error of a {\em single} measurement of position angle, not the mean error in the determination of $\delta \phi$ from measurements at multiple positions.

From SCUBA $\lambda850$ $\mu$m maps of a number of cores, including the three discussed in this paper, \citet{k03} and \citet{kwa03} have inferred the core masses. Our estimate of the field strength in each core will apply to the region over which we map the polarization; we therefore use mean values for parameters needed in equations (1) and (2) appropriate to the area of the polarization measurements. For L1544 and L43 the polarization detections are confined to positions with intensities of about half the peak intensity or higher, so we use the inferred mass within the half-power intensity contour. However, for L183 our polarization map covers an area significantly larger than the half-power contour. Although there are four measurements outside the 100 mJy beam$^{-1}$ contour, almost all measurements are within this intensity contour. We therefore use the mass within this contour to infer needed parameters. With the approximation that the mean radius $\bar{r}$ is the geometric mean of the major and minor axes of the selected intensity contour, we estimate $N(H_2) \approx M/\pi\bar{r}^2m$ and $n(H_2) \approx 3M/4\pi\bar{r}^3m$ from the SCUBA masses and mean radii; we use the distances to each core from \citet{k03} and \citet{kwa03}. 

We use measurements of the N$_2$H$^+$ (1-0) line \citep{cbmt02} to obtain $\Delta V$. There are several reasons to support this being the best available choice. The excitation of the N$_2$H$^+$ (1-0) line suggests that it traces high-density material similar to that traced by the dust emission. The \citet{cbmt02} spectral-line maps of L183, L1544, and L43 agree well with the SCUBA dust maps, except that the deconvolved (corrected for telescope beam broadening) sizes of the cores in N$_2$H$^+$ are about 50\% larger than in dust emission sizes. This of course suggests that N$_2$H$^+$ samples on average a slightly lower density region than does the $\lambda850$ $\mu$m dust emission. In each subsection below we compare $n(H_2)$ inferred from N$_2$H$^+$ excitation with values from dust emission for each core. The expectation that N$_2$H$^+$ indeed samples a slightly lower density is borne out by these comparisons; we use the densities inferred from the dust emission results in equation (2). However, we need estimates of the velocity dispersions from spectral-line data. We use the values of $\Delta V$(N$_2$H$^+$) at each peak position as the best available value. Use of the peak value rather than a value averaged over a larger region perhaps compensates for the fact that N$_2$H$^+$ samples slightly lower densities. Moreover, although there is some structure in $\Delta V$(N$_2$H$^+$) shown in the maps of \citet{cbmt02}, over the roughly $2^\prime \times 1^\prime$ area of each of our three cores this variation is small. In addition to the single-dish data, we make use of the high-resolution BIMA maps of L1544 in the N$_2$H$^+$ (1-0) line \citep{wmwf99}. Finally, since we need the turbulent velocity dispersion in equation (2), we correct each total N$_2$H$^+$ line width for the (small) thermal contribution to the line width (with $T_K = 10$ K) and use $\Delta V_{NT}$ in equation (2).

Table 1 lists the relevant data and the results, where $D$ is the distance of each core, $\bar{r}$ is the mean core radius on the plane of the sky (the half-power radius except for L183), $\Delta V_{NT}$ is the turbulent FWHM line width, $M_{dust}$ is the total (gas and dust) core mass inferred from the dust emission within the $\lambda 850$ $\mu$m dust emission  contour selected for each core, $N(H_2)$ and $n(H_2)$ are the mean column and volume densities inferred from $M_{dust}$ and $\bar{r}$, $\delta \phi$ is the corrected polarization position angle dispersion, $B_{pos}$ is the field strength in the plane of the sky, $\lambda_{obs}$ is the observed mass-to-magnetic flux ratio in units of the critical ratio, $\lambda_{cor}$ is $\lambda_{obs}$ corrected by 1/3 from equation (4), $M_{B crit}$ is the magnetic critical mass that can be supported by our measured $B$, and $M_{vir}$ is the virial mass (e.g., \citep{cbmt02}), computed for $T_K = 10$ K and the $\bar{r}$ and $\Delta V_{NT}$ listed in the table. In each subsection below, we discuss each core separately. 

\subsection{L183}

The L183 core (Figure 1) is only slightly elongated at the JCMT resolution, with a position angle of $\sim -15^\circ$; this core is embedded in a filament-like structure extending further north and south than our SCUBA map. The filament has approximately the same position angle south of the core as does the core itself, but north of the core it is nearly north-south. On the eastern side of the filament, polarization is detected at a threshold total intensity level of about the 60 mJy beam$^{-1}$. On the western side the polarization detection threshold is about a factor of two higher. 

There are 25 independent positions from which the mean polarization position angle and the dispersion about that mean in the position angles may be calculated. The signal-to-noise weighted mean position angle is $\bar{\phi} = 28^\circ \pm 2^\circ$. The signal-to-noise weighted dispersion is $\delta \phi = 15^\circ \pm 2^\circ$, while the contribution to $\delta \phi$ due to measurement uncertainty is $6^\circ$. Corrected for the measurement uncertainty contribution, $\delta \phi = 14^\circ$. 

However, one must examine the possibility that there is a contribution to this dispersion due to a large-scale change in the field direction over the core. If one ignored this, the calculated dispersion would be too large, and equation (2) would underestimate $B_{pos}$. To check for a variation in the mean field direction due to the bend in the filament, we computed the mean position angle north ($\bar{\phi}_N$) and south ($\bar{\phi}_S$) of $\delta = -2^\circ 44^\prime$, the declination of the bend. Results were $\bar{\phi}_N = 29^\circ \pm 2^\circ$ and $\bar{\phi}_S = 25^\circ \pm 2^\circ$; therefore, there is no significant systematic change in the mean field direction due to the bend in the filament.  The position angles do appear to be more random in the south than the north, however; $\delta \phi_N = 11^\circ \pm 2^\circ$ while $\delta \phi_S = 17^\circ \pm 2^\circ$. One possible explanation is that there is a south-to-north gradient in $B_{pos}$. However, a gradient in $n(H_2)$ and/or $\delta V$ and a constant $B_{pos}$ would produce the same effect. There is no gradient observed in $\Delta V$ (see \citet{cbmt02} Figure 8) across the core. We can also carry out this experiment by dividing the data into east and west regions. We computed the mean position angle east ($\bar{\phi}_E$) and west ($\bar{\phi}_W$) of $\alpha = 15^h 51^m 32^s$. Results were $\bar{\phi}_E = 31^\circ \pm 3^\circ$ and $\bar{\phi}_W = 19^\circ \pm 4^\circ$; therefore, there may be a small difference in the mean field direction east and west of the core. However, this does not significantly affect the dispersion in position angles. Computing the position angle dispersions separately in the two regions yields $\delta \phi_E = 14^\circ \pm 3^\circ$ and $\delta \phi_W = 14^\circ \pm 4^\circ$; therefore, there is little change from the result ($\delta \phi = 15^\circ$) obtained above assuming that the mean field is uniform in direction over the entire core. 

It seems clear that the polarization data do not sample the innermost core region, for there is depolarization toward the inner core ($r < 15^{\prime\prime}$ from the peak intensity position) in comparison with the envelope of the core. Such polarization ``holes'' are common (but not ubiquitous). \citet{mwf01} argued that such depolarization seen in the OMC3 molecular filament was due to geometry -- the effect of a projected helical magnetic field. However, the depolarization toward L183 is not seen along the filament as was the case toward OMC3, but rather is confined to the inner core. Depolarization is also seen quite strongly in dense cores mapped interferometrically (e.g., \citet{l02}). A possible explanation is a change in grain properties such that grains at high densities are not polarizing. Polarization maps usually present percentage polarization rather than the polarized intensity, PI. In Figure 1 we have plotted PI in order to show that it is not just the case that the percentage polarization decreases in the inner core, but that the PI itself is weaker toward the core. PI is fairly uniform across the source outside of the inner core. The polarization ``hole'' (or conversely, polarization limb brightening) suggests that the polarization arises predominantly in an envelope or shell region. Toward the center positions the shorter line of sight through the envelope, with the inner core itself producing no polarization, could explain the observations. 

We measure polarization at almost all positions within the 100 mJy beam$^{-1}$ contour (and a few postions to the southeast outside this contour). We therefore apply the Chandrasekhar-Fermi relation over the region within the 100 mJy beam$^{-1}$ contour. From the SCUBA data we infer a mass within the 100 mJy beam$^{-1}$ contour to be 1.0 M$_\odot$ and $\bar{r} \approx 0.023$ pc, which yield a mean density $\bar{n}(H_2) \approx 2.9 \times 10^5$ cm$^{-3}$. If we assume that the inner core does not contribute to the polarization, we must exclude its contribution to the mass. The SCUBA data yield a mass of 0.24 M$_\odot$ within the cylinder with a radius of 15$^{\prime\prime}$ (0.008 pc) centered on the peak position. Subtracting this mass and the volume of the cylinder from the total mass and volume, we find $\bar{n}(H_2) \approx 2.6 \times 10^5$ cm$^{-3}$. Hence, the assumption that polarization does not sample the inner core makes little difference in the estimate of $\bar{n}(H_2)$ and hence of the magnetic field strength. We find $N(H_2) \approx 2.7 \times 10^{22}$ cm$^{-2}$ for the envelope region. 

We use the lower $n(H_2)$ density appropriate to the envelope, but note that a two-level excitation analysis of the N$_2$H$^+$ (1-0) line \citet{cbmt02} found $\bar{n}(H_2) \approx 0.8 \times 10^5$ cm$^{-3}$. As discussed above, the N$_2$H$^+$ (1-0) line appears to sample slightly lower density than does the dust emission.

The N$_2$H$^+$ line width at the peak position is $\Delta V = 0.25$ km s$^{-1}$. Correcting for the thermal contribution, we have $\Delta V_{NT} = 0.22$ km s$^{-1}$ for the turbulent line width. Thus, with $n(H_2) = 2.6 \times 10^{5}$ cm$^{-3}$, $\Delta V = 0.22$ km s$^{-1}$, and $\delta \phi = 14^\circ$, equation (2) yields $B_{pos} \approx 80$~$\mu$G.

\citet{c93} carried out Zeeman observations in the 18-cm lines of OH toward dark clouds, including L183. They set a $3-\sigma$ upper limit for the mean line-of-sight field strength of $B_{los} < 16$ $\mu$G, which is much smaller than $B_{pos} \approx 80$ $\mu$G inferred here. It is possible that {\bf B} lies very nearly in the plane of the sky. However, a more likely explanation for the difference is that the OH Zeeman observations did not sample the dust core. First, quasi-thermal OH emission almost certainly does not sample regions with $n(H_2) > 10^4$ cm$^{-3}$. Moreover, the OH observations were carried out with an $18^\prime$ beam, so that an area of sky about two orders of magnitude larger was sampled for $B_{los}$ than for $B_{pos}$. Indeed, the mean (over the $18^\prime$ beam) $n(H_2)$ inferred from the OH data is $n(H_2) \approx 1.3 \times 10^3$ cm$^{-3}$, about two orders of magnitude smaller than the density in the core region sampled by the dust and N$_2$H$^+$ line emission. For constant mass-to-magnetic flux ratio, $B$ scales as $\sqrt{\rho}$, so it is not surprising that the upper limit on $B$ inferred from the OH Zeeman data is significantly smaller than $B$ inferred from the dust polarimetry.

For $N(H_2) = 2.7 \times 10^{22}$ cm$^{-2}$ and B$_{pos} = 80$ $\mu$G, we have $\lambda \approx 2.6$, or slightly supercritical. Hence, the regular magnetic field alone is not sufficient to support the core against gravity. Without support from other sources, the L183 core should be collapsing. In fact, in a survey for infall motions, \citet{lmt01} found that L183 showed infall motions, consistent with collapse. However, if we apply the statistical correction factor from equation (4), $\lambda_{cor} \approx 0.9$.

\subsection{L1544}

No new SCUBA data for L1544 were obtained; here we analyze the data of \citet{dwt00} -- see their Figure 1. Toward L1544 there were only 9 independent positions where polarization was measured. All positions were within the 60\% of peak contour. Although the percentage polarization inside the 80\% of peak contour was lower than at positions farther from the center, there were no significant polarization detections south of the core. Hence, the strength of the polarization appears to decrease from north to south without a clear inner core depolarization. The signal-to-noise weighted dispersion in the polarization position angles is $\delta \phi = 15^\circ \pm 2^\circ$, while the contribution to $\delta \phi$ due to measurement uncertainty is $6^\circ$. Corrected for the measurement contribution, $\delta \phi = 13^\circ$. 

From the SCUBA data we infer the mean density within the half-power contour $\bar{n}(H_2) \approx 4.9 \times 10^5$ cm$^{-3}$. From an excitation analysis of the N$_2$H$^+$ (1-0) line \citet{cbmt02} find $\bar{n}(H_2) \approx 0.7 \times 10^5$ cm$^{-3}$, which is significantly smaller than the SCUBA value. Although we noted above that the N$_2$H$^+$ (1-0) line appears to sample lower densities than the dust emission, there may be another reason for this rather large difference in L1544. \citet{Taf98} and \citet{wmwf99} studied L1544 in detail and suggested that the observed redshifted self-absorption in spectral lines implied that L1544 has a low-density envelope with infall onto a high-density core. \citet{wmwf99} observed the N$_2$H$^+$ (1-0) line at high angular resolution with BIMA and carried out a detailed analysis. Their model has two components, a foreground low density region with $n(H_2)_{peak} \approx 0.1 \times 10^5$ cm$^{-3}$ and a background dense region with $n(H_2)_{peak} \approx 4 \times 10^5$ cm$^{-3}$. The single-dish observations of \citet{cbmt02} were analyzed with only a single component, and they found a density intermediate between those of the two density components. The dust emission mapped with SCUBA is optically thin and requires a large column density for detection; it would not be sensitive to a low-density foreground envelope. Hence, the density of the core inferred from the BIMA data and from the SCUBA data agree rather well. We use the density inferred from the SCUBA data. 

The N$_2$H$^+$ line width at the peak position is $\Delta V = 0.31$ km s$^{-1}$. Correcting for the thermal contribution, we have $\Delta V_{NT} = 0.28$ km s$^{-1}$ for the turbulent line width. Thus, for $n(H_2) = 4.9 \times 10^{5}$ cm$^{-3}$, $\Delta V_{NT} = 0.28$ km s$^{-1}$, and $\delta \phi = 13^\circ$, equation (2) yields $B_{pos} \approx 140$~$\mu$G. 

\citet{ct00} detected the Zeeman effect in the 18-cm lines of OH with the Arecibo telescope and inferred $B_{los} \approx 11$ $\mu$G, a much smaller value than our estimate here of the field strength in the plane of the sky. However, \citet{ct00} argued that the OH data did not sample the small dense core observed in $\lambda850$ $\mu$m dust emission. Part of the problem is that the $3^\prime$ angular resolution of the Arecibo telescope is about 5 times the area of the SCUBA dust core. Probably more importantly, \citet{ct00} argued that OH and the C$^{18}$O (1-0) lines sample the same gas, and higher angular resolution ($46^{\prime\prime}$) C$^{18}$O (1-0) maps \citep{Taf98} show no evidence for the small, dense dust core. Rather, the C$^{18}$O (and OH) lines seem to sample a lower density envelope region, perhaps the foreground component with $n(H_2) \approx 0.1 \times 10^5$ cm$^{-3}$ discussed by \citet{wmwf99}. The flux-freezing scaling $B \propto \sqrt{\rho}$ would account for the difference in $B_{pos}$ and $B_{los}$ without requiring that the field be mainly in the plane of the sky.

For $N(H_2) = 4.2 \times 10^{22}$ cm$^{-2}$ and B = 140 $\mu$G, we have $\lambda \approx 2.3$, or slightly supercritical. Hence, the regular magnetic field alone is insufficient to support the core against gravity. This result would be consistent with the previous claims that L1544 is infalling (collapsing). Again, however, application of the possible geometrical correction from equation (4) would yield $\lambda_{cor} \approx 0.8$.

\subsection{L43}

For L43 no new SCUBA data were obtained either, and we analyze the data of \citet{dwt00} -- see their Figure 3. Toward L43 there were 31 independent positions where the position angle of the polarization was measured. However, L43 is a considerably more complicated cloud than the other two. The prestellar core we are interested in is not isolated; there is a second core at the western edge of the SCUBA field of view \citep{dwt00} in which a classical T Tauri star, RNO 91 \citep{c80}, is embedded. The outflow from RNO 91 has cleared a cavity to the south of the source and the southern edge of the prestellar core forms part of the cavity wall. The direction of the SCUBA polarization vectors from the center of the prestellar core to the north do not appear to have been affected by the cavity. However, to the south and west the position angles are distinctly different, with the magnetic field turning smoothly through an angle of roughly $90^\circ$ from the prestellar core position to the RNO 91 position. A blind calculation of the dispersion of the position angles over the entire field would include a very significant contribution from this structure in the regular field. Because we are interested in the field in the prestellar core, we use only those positions on and north of the core that appear to be unaffected by RNO 91; specifically, those north of $\delta(1950) = -15^\circ 41^\prime$ and east of $\alpha(1950) = 16^h 31^m 42^s$. This leaves 11 positions that sample the magnetic field in the L43 prestellar core. All positions are within the 50\% of peak contour. The strength of the polarization appears to decrease from west to east with only a weak inner core central depolarization. The signal-to-noise weighted dispersion in the polarization position angles for the 11 core positions is $\delta \phi = 13^\circ \pm 1^\circ$, while the contribution to $\delta \phi$ due to measurement uncertainty is $4^\circ$. Corrected for the measurement contribution, $\delta \phi = 12^\circ$. 

From the SCUBA data we infer the mean density within the half-power contour $\bar{n}(H_2) \approx 3.8 \times 10^5$ cm$^{-3}$. From an excitation analysis of the N$_2$H$^+$ (1-0) line \citet{cbmt02} find $\bar{n}(H_2) \approx 2.2 \times 10^5$ cm$^{-3}$, which agrees  well with the SCUBA result, especially considering the evidence discussed above that the N$_2$H$^+$ (1-0) line may sample slightly lower densities than the dust emission. Their N$_2$H$^+$ (1-0) line width is $\Delta V = 0.36$ km s$^{-1}$. Correcting for the thermal contribution, we have $\Delta V_{NT} = 0.34$ km s$^{-1}$ for the turbulent line width. For $n(H_2) = 3.8 \times 10^{5}$ cm$^{-3}$, $\Delta V_{NT} = 0.34$ km s$^{-1}$, and $\delta \phi = 12^\circ$, equation (2) yields $B_{pos} \approx 160$ $\mu$G. There are no Zeeman data on the line-of-sight component of the field toward L43.

For $N(H_2) = 3.9 \times 10^{22}$ cm$^{-2}$ and B = 160 $\mu$G, we have $\lambda \approx 1.9$, or slightly supercritical. Hence, as for the other two cores, the regular magnetic field alone is insufficient to support the core against gravity. There are no data available to address whether or not infall motions are seen toward the L43 prestellar core, but it is a part of a cloud in which a T Tauri star has formed nearby. The possible geometrical correction from equation (4) would yield $\lambda_{cor} \approx 0.6$.

\section{Conclusion}

We have applied the Chandrasekhar-Fermi method to estimate magnetic field strengths in the plane of the sky in three dense, prestellar cores. Our major result inferred directly from the observations is that for all three cores the inferred mass-to-magnetic flux ratios are supercritical by approximately a factor of two, or $\lambda \approx 2$. Said another way, the mass that can be supported by the measured plane-of-sky field, $M_{Bcrit}$, is about 1/2 the actual mass, $M_{dust}$, inferred from the intensity of the dust emission. Hence, the ordered magnetic fields in these cores are insufficient to support the cores against gravitational collapse. 

However, cores which are supported by magnetic fields would have a disk morphology with the field along the minor axis. In the extreme case of plane-parallel disk geometry, the statistical mean correction would be $\lambda_{cor} = \frac{1}{3}\lambda_{obs}$. Although this correction is a statistical one that is only valid when considering the mean value of $\lambda$ for a large number of observations, applying it to the three cores discussed here changes the formal results such that $\overline{\lambda_{cor}}$ is critical to slightly subcritical. But of course, the uncertainties in inferring $\lambda$ are such that it is not possible to say whether $\lambda$ in these cores is slightly subcritical or slightly supercritical.

Surprisingly, the virial masses are essentially equal to the observed dust masses. The fact that  $M_{vir} \approx M_{dust}$ suggests that the cores are supported by turbulent and thermal motions (mainly thermal, since $\Delta V_{th} = 0.44$ km s$^{-1}$ for a mean particle mass of 2.33m$_H$ and T = 10 K). However, for L183 and L1544 infall motions have been inferred from self-absorbed molecular line profiles, suggesting collapse; data on possible infall are not yet available for L43. It seems likely that the virial masses are overestimated. \citet{cbmt02} found a similar inconsistency in their data; {\em all} their cores with $M_{vir} < 2.5 M_\odot$ have $M_{vir} \approx 2M_{ex}$, where $M_{ex}$ was calculated with the density inferred from collisional excitation of the N$_2$H$^+$ line. 

There are several factors that would lead to virial masses being overestimated. First, virial masses are based on a crude model --  a uniform density sphere. \citet{cbmt02} pointed out that a cloud model with a centrally peaked density distribution with $\rho \propto r^{-2}$ would change the virial formula such that $M_{vir}$ would be reduced by a factor of 1.6. The radial density profiles of starless cores are observed to be centrally peaked, based on analysis of the data with isothermal models. However, the central peaking may be even more pronounced than the isothermal models imply. Modelling of dust emission observations have led to evidence for temperature decreasing inward (\citep{zwg01} and \citep{e01}. Since the intensity of the dust emission is approximately proportional to the product of the column density and the temperature, an outward temperature gradient would require a more peaked density distribution to fit the observations than the isothermal fits have yielded. In addition, kinetic temperatures within the core radii over which polarization is measured would be smaller than the mean; so the $\Delta V_{th}$ used in the virial mass equation would be lowered, further reducing the virial mass. Hence, the inconsistency would be resolved, and the virial masses would be significantly below the observed masses. With $M_{Bcrit} + M_{vir} < M_{dust}$, collapse would be possible. These changes would also affect the $B_{pos}$ and $\lambda$ that we infer, but insufficiently detailed information on temperatures is available to justify trying to take this into account at this time.

Can our results help distinguish between the ambipolar-diffusion and turbulence driven models of star formation? The magnetically supported model of star formation does make a specific prediction. For an initially subcritical cloud, ambipolar diffusion increases the mass-to-flux ratio in the core until the core becomes supercritical. In a typical model, after the central density $\rho_c$ has increased by $\sim 10^{1.5}$ from its initial value $\rho_0$ and the core becomes magnetically critical, the mass-to-flux ratio in the core will slowly increase from $\lambda = 1$ to $\lambda \approx 3$ as $\rho_c/\rho_0$ increases to 10$^6$ (e.g., \citet{cm94}, Figure 4). Hence, in spite of the uncertainties in inferring $\lambda$, this prediction appears to be well satisfied by our data. 

However, in spite of the fact that the observations satisfy the ambipolar-diffusion driven prediction, this does not rule out turbulence-driven star formation. The turbulence models do not make a specific prediction, except of course that $\lambda > 1$. If $\lambda \sim 2$ in the general, turbulent interstellar medium, those values of $\lambda$ would persist when cores were formed by turbulence, so long as flux freezing held. Hence, while the result $\lambda \approx 2$ is suggestive, it is not definitive. While measurements of $\lambda$ could rule out a dominant role for magnetic fields if $\lambda >> 1$ were found in cores, they could only rule out a dominant role for turbulence if subcritical mass-to-flux ratios were observed at earlier stages of cloud evolution, before dense cores formed.

This analysis has shown that dust polarization mapping of dense prestellar cores can provide information about magnetic field strengths. Especially in the absence of Zeeman observations that sample these dense regions, this technique adds significantly to our extremely sparse knowledge about the role of magnetic fields in the star formation process. 

\acknowledgments This work was partially supported by NSF grant AST 02-05810. DJN wishes to thank PPARC for studentship support. 

\clearpage


\clearpage

\figcaption {Dust continuum emission at $\lambda850$ $\mu$m from the L183 prestellar core. Coordinates are B1950. The Stokes I map is shown as a gray scale with contours overlaid. Contour levels are at 50, 100, 150, 200, 250, and 300 mJy beam$^{-1}$; the peak flux density is 333 mJy beam$^{-1}$. The direction of the B-field in the plane of the sky is shown by a series of lines at every position where a measurement of the polarized flux above the $2-\sigma$ level was achieved. The plotted B-vectors are perpendicular to the direction of the polarization observed. The length of each B-vector is proportional to the polarized intensity; the scale is 1 mJy beam$^{-1} = 1^{\prime\prime}$. The vectors are plotted on a $21^{\prime\prime}$ grid spacing, equal to the spatial resolution after smoothing, so each vector is independent.} 

\begin{figure} 
\plotone{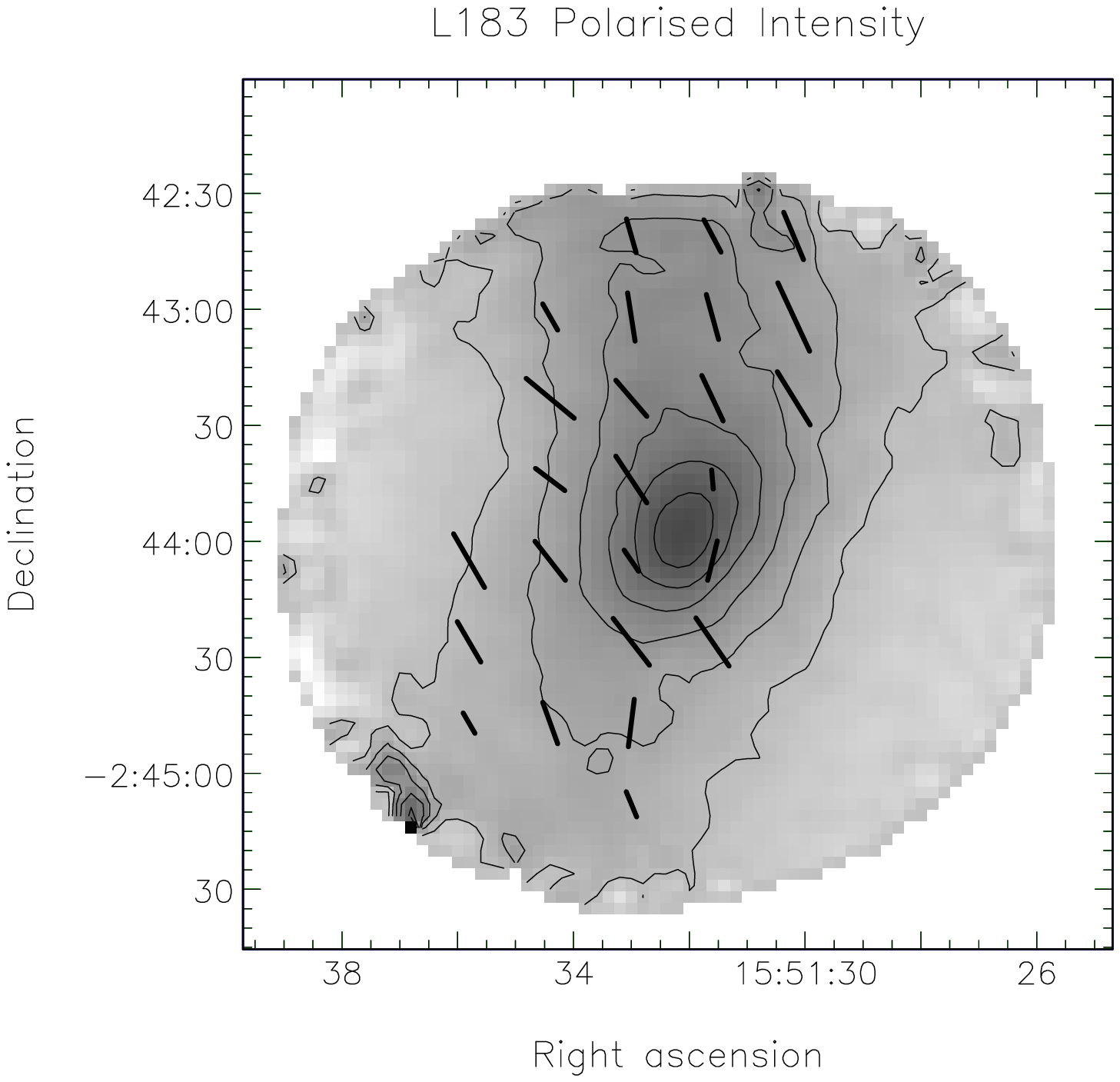}

\end{figure}
\clearpage

\begin{deluxetable}{lccc}
\tablecaption{Core Data and Magnetic Field Strengths \label{tbl-2}}
\tablewidth{0pt}
\tablehead{\colhead{} & \colhead{L183} & \colhead{L1544} & \colhead{L43} }
\startdata
$D(pc)$ & 110 & 140 & 130\\
$\bar{r}$(pc) & 0.023 & 0.021 & 0.025\\
$\Delta$V$_{NT}$(km s$^{-1}$) & 0.22 & 0.28 & 0.34\\
$M_{dust}(M_\odot$) & 1.0 & 1.3 & 1.7\\
$N_{H_2}$($10^{22}$ cm$^{-2}$) & 2.7 & 4.2 & 3.9\\
$n_{H_2}$($10^5$ cm$^{-3}$) & 2.9 & 4.9 & 3.8\\
$\delta\phi(^\circ)$ & 14 & 13 & 12\\
$B_{pos}$($\mu$G) & 80 & 140 & 160\\
$\lambda_{obs}$ & 2.6 & 2.3 & 1.9\\
$\lambda_{cor}$ & 0.9 & 0.8 & 0.6\\
$M_{B crit}(M_\odot$) & 0.4 & 0.6 & 0.9\\
$M_{vir}(M_\odot$) & 1.1 & 1.1 & 1.6\\
\enddata
\end{deluxetable}

\end{document}